\documentstyle[prl,aps,preprint,epsf]{revtex}

\title{Correlations and fluctuations of a confined electron gas}
\author{P. Leboeuf and A. Monastra}
\address{Laboratoire de Physique Th\'eorique et Mod\`eles Statistiques
\footnote{Unit\'e de recherche de l'Universit\'e de Paris XI associ\'ee au 
CNRS}, B\^at. 100, \\ 91405 Orsay Cedex, France}

\begin{document}
\maketitle {\begin{abstract} The grand potential $\Omega$ and the response $R
= - \partial \Omega /\partial x$ of a phase-coherent confined noninteracting
electron gas depend sensitively on chemical potential $\mu$ or external
parameter $x$. We compute their autocorrelation as a function of $\mu$, $x$
and temperature. The result is related to the short-time dynamics of the
corresponding classical system, implying in general the absence of a universal
regime. Chaotic, diffusive and integrable motions are investigated, and
illustrated numerically. The autocorrelation of the persistent current of a
disordered mesoscopic ring is also computed.
\end{abstract}}

\vspace{2.5cm}


\vspace{2.5cm}

\pagebreak

\narrowtext
A major advance in mesoscopic physics has been the observation of
single-electron interference or more generally of quantum mechanical effects
in mesoscopic objects. Among these we can mention the supershell structure
in atomic clusters, the persistent currents in mesoscopic rings, or
the quantization of the longitudinal conductance in microjunctions
\cite{houches2}.

In all these experiments the spatial confinement of the electron gas plays a
significant role. When a parameter controlling the system is varied -- like
the chemical potential, a magnetic field, or the shape of the confining potential -- the response of the gas shows irregular fluctuations as a
function of the parameter. While the average value of the variation is related
to a macroscopic property of the system (like its volume), the wild
fluctuations are the imprint of quantum mechanical interference.

Neglecting the interactions between the electrons, our purpose is to provide a
general description of the fluctuating properties of the many-particle ground
state electron gas when different parameters -- like temperature $T$, chemical
potential $\mu$, or external parameter $x$ -- are varied. Taking as
reference parameters $(x_0,\mu_0,T_0)$, the system is described in the
grand-canonical ensemble by the grand potential
\begin{equation} 
\Omega (x_0,\mu_0,T_0)= - \int dE ~ {\cal N} (E,x_0) ~ f(E,\mu_0,T_0) \ , 
\label{grand}
\end{equation} 
with ${\cal N} (E,x_0) = g_s \sum _k \Theta ( E - E_k(x_0) )$ the counting
function of the single particle states $E_k(x_0)$, $\Theta$ the Heavyside step
function, $f(E,\mu_0,T_0)= 1/[1 + exp ( (E - \mu_0)/k T_0 ) ]$ the Fermi
distribution and $g_s=2$ for spin degeneracy.

The reaction of the system to a variation of the external parameter is 
characterized by the response function
$$
R(x_0,\mu_0,T_0)= - \left. \partial \Omega/\partial x_0 \right|_{\mu_0,T_0} \ . \label{r}
$$
The physical interpretation of $R$ depends on the geometry of the sample and
the nature of the parameter $x$. It is a magnetic moment if $x$ is a magnetic
field, a persistent current when the sample geometry is annular and
$x$ is an Aharonov-Bohm flux, or a force when $x$ controls the shape of the
confining potential (like when deforming a metallic cluster for example).

We characterize the fluctuating properties of the electron gas by computing
the autocorrelation function
\begin{equation}
C_{\Omega}(x,\mu,T) = \left\langle \widetilde{\Omega} ({\bf q}_{-}) 
~ \widetilde{\Omega}({\bf q}_{+}) \right\rangle_{x_0,\mu_0} \ , \label{corr1}
\end{equation}
with ${\bf q}_{\pm} = (x_0\pm x/2,\mu_0\pm \mu/2,T_0\pm T/2)$. An analogous
definition holds for $C_R (x,\mu,T)$. We denote the fluctuating part by an
upper tilde symbol (see below). The brackets denote some energy and parameter
averages.

Parametric correlations like $C_{\Omega}$ or $C_R$ have been studied in the
past for chaotic and disordered systems for ``individual'' (single-particle)
levels, rather than for a many-body ground state. The single-particle analog
of $C_R(x,0,0)$ is the velocity-velocity correlation \cite{sza}. Although for
the latter no global expression is known, its tail $x\rightarrow\infty$ has
been shown to be universal \cite{sza,sa1,been}. In contrast to this, we will
see below that for a Fermi gas the parametric correlators are explicitly
computable and are expressed semiclassically, even at zero temperature, in
terms of the classical short time dynamics of the system. Because the latter
is system-specific, the correlators are in general system-dependent with no
universal regime whatsoever. This difference with the universal local
properties occurs mainly because the typical time scale of a perturbed ground
state Fermi gas is $h/\mu_0$, which is much smaller than the Heisenberg
time.

We express the counting function in Eq.(\ref{grand}) as a sum of smooth plus
oscillatory terms ${\cal N} = \overline{{\cal N}} + \widetilde{{\cal N}} $.
The smooth part is given by the usual Weyl or Thomas-Fermi approximation.
Semiclassically, the oscillatory part is $\widetilde{{\cal N}} = 2 \hbar g_s
\sum_p (A_p /\tau_p) \sin(S_p/\hbar)$. The sum is over the periodic orbits $p$
with action $S_p$ (including the Maslov index), period $\tau_p=\partial
S_p/\partial E$, and stability amplitude $A_p$ \cite{gutz}. This sum is
inserted in Eq.(\ref{grand}) to compute the oscillatory part of the grand
potential. To leading order in $\hbar$ and for $k T_0 \ll \mu_0$ \cite{sm,ruj},
\begin{equation}
\widetilde{\Omega} (x_0,\mu_0,T_0) \approx 2 \hbar^2 g_s \sum_p 
\widetilde{A}_p \cos(S_p/\hbar) \label{grandosc} \ ,
\end{equation}
with $\widetilde{A}_p = A_p \ \kappa(\tau_p,T_0)/\tau_p^2$. The factor
$\kappa(\tau_p,T_0) = \frac{\tau_p / \tau_c}{\sinh (\tau_p / \tau_c)}$, with
$\tau_c = \hbar/ \pi k T_0$ takes into account finite temperature effects and
acts as an exponential cut off for long periodic orbits. All the classical
quantities entering this expression are evaluated at $(x_0,\mu_0)$. The main
dependence on $x$ comes from the oscillatory terms via the dependence of the
actions, $S_p = S_p(x_0,\mu_0)$. Taking this into account, the oscillating part
of the response function $\widetilde{R} = - \partial
\widetilde{\Omega}/\partial x_0|_{\mu_0,T_0}$ is
\begin{equation}
\widetilde{R} (x_0,\mu_0,T_0) \approx 2 \hbar g_s \sum_p \widetilde{A}_p ~ Q_p
\sin(S_p/\hbar) \ , \label{rosc}
\end{equation}
with $Q_p = \partial S_p/\partial x_0 |_{\mu_0,T_0}$. 

Replacing Eq.(\ref{grandosc}) in (\ref{corr1}) we obtain a double sum over
periodic orbits with oscillations depending on the difference of their
actions. The averages in Eq.(\ref{corr1}) are done over some range $\Delta
\mu$ and $\Delta x$ which contain several oscillations of $\widetilde{\Omega}$
but are on the other hand small in comparison to the scale over which the
classical functions vary. As mentioned before, the main dependence on
$(x,\mu)$ comes from the actions, which may be linearized with respect to
these parameters. The average restricts the sum to orbits having approximately
the same $Q_p$ and $\tau_p$. Ordering the orbits by their period, taking into
account these restrictions and using the semiclassical definition of the
parametric form factor $K(\tau,x)$ (i.e., the Fourier transform of the
two-point correlation function \cite{sa1}) we obtain
\begin{equation}
C_{\Omega} = \frac{\hbar^2 g_s^2}{2 \pi^2} \int_0^{\infty}
\frac{d\tau}{\tau^4}~ \chi (\tau,T) ~ \cos ( \mu \tau /\hbar ) ~
K(\tau,x) \ , \label{ck}
\end{equation}
with $\chi (\tau,T) = \kappa (\tau,T_0-T/2)\ \kappa (\tau,T_0+T/2)$. The
autocorrelation of the response function follows from Eq.(\ref{ck}) by
derivation with respect to the external parameter.

Due to the factor $\tau^4$ in the denominator, the integral in (\ref{ck}) is
dominated by the short-time dynamics of the system. For chaotic systems a
naive replacement of the form factor $K(\tau,x)$ by the random matrix result
produces a divergence. In real systems, this divergence is avoided by the cut
off introduced by the shortest periodic orbit of period $\tau_{min}$ (the form
factor vanishes for $\tau<\tau_{min}$). Then even at $T_0 = T = 0$ ($\chi=1$),
the autocorrelation will be described with good accuracy by computing in
Eq.(\ref{ck}) only the contribution of the shortest periodic orbits, the
longer ones representing a small correction. Because this short-time dynamics
is system-specific, then in general we do not expect any universality for
$C_\Omega$ and $C_R$.

Eq.(\ref{ck}) may therefore be restricted to the diagonal approximation
$p=p'$ for the form factor, $K_D (\tau)= h^2 \sum_{p} A_p^2 \ \delta
(\tau-\tau_p)$, with the result
\begin{equation}
C_{\Omega} = \frac{\hbar^2 g_s^2}{2 \pi^2} \int_0^{\infty}
\frac{d\tau}{\tau^4} \chi \cos ( \mu \tau /\hbar )
\langle \cos ( Q x /\hbar ) \rangle K_D (\tau) \ .
\label{co2}
\end{equation}
Here the average is evaluated over the $Q_p$'s taken among all the periodic
orbits having a period between $\tau$ and $\tau+d\tau$. The expression for the
response function is
\begin{equation}
C_R = \frac{g_s^2}{2 \pi^2} \int_0^{\infty} \frac{d\tau}
{\tau^4} \chi \cos ( \mu \tau /\hbar ) \langle Q^2 
\cos ( Q x /\hbar ) \rangle K_D (\tau) \ .
\label{cr2}
\end{equation}
To leading order, Eq.(\ref{co2}) is also valid for systems with a fixed number
of particles $n$ \cite{agi,ruj}. The form factor is now evaluated at $E=
\bar{\mu} (n,x_0)$ obtained by inverting $n=\overline{\cal N}
(\bar{\mu},x_0)$. On the other hand, Eq.(\ref{cr2}) is modified according to
$Q^2 \rightarrow Q^2 + \tau^2 \left ( \partial \bar{\mu}/\partial x_0
\right ) ^2$. The importance of this new term depends on the nature of the
external parameter. If $x$ is a flux in an Aharonov-Bohm geometry,
$\overline{\cal N}$ is independent of $x$ and $\partial \bar{\mu}/ \partial
x_0 =0$. On the other hand, if $x$ is the deformation of a two dimensional
billiard, we find that the contribution of this additional term is comparable
to the $Q^2$ term if the area of the billiard varies, while it is smaller by a
factor $1/n$ when the area is kept fixed.

Eqs.(\ref{co2}) and (\ref{cr2}) are the basic results of this paper. We now
proceed to analyze them for different physical situations. First we consider a
ballistic motion in a regular cavity (the trajectories of the electrons occur
on phase-space tori). As an example of regular dynamics we have studied a
Fermi gas contained in a two dimensional rectangular box of fixed area ${\cal
A}$ and sides $a$ and $b$. We choose $x=a/b$ as the external parameter. Exact
expressions for $A_p,\tau_p, S_p$ and $Q_p$ are known in this case. We can
therefore explicitly evaluate $C_\Omega$ and $C_R$ as a function of $\mu$ or
$x$. In Fig.~1 we show the numerical results for $x_0 \approx 1.3$ compared to
the theoretical curves using only the first $10$ orbits as a function of the
rescaled variable $\xi = \sqrt{\mu_0} ~ x/x_0^{3/2}$. We observe erratic
long-range oscillations accurately described by our formulas.

We now turn our attention to chaotic systems. Considering that the form factor
is strictly zero for $\tau < \tau_{min}$, the simplest approximation one can
made for these systems is to assume a random matrix theory \cite{mehta}
behavior for $K_D (\tau)$ starting at $\tau = \tau_{min}$
\begin{equation} \label{kbmin}
K_D (\tau) = \left\{ \begin{array}{ll} 
0  \ , \;\;\;\;\;\;\; & \tau < \tau_{min} \\ 
2 \tau/\beta \ , \;\;\;\; & \tau_{min} < \tau << \tau_H \ . \end{array} \right.
\end{equation}
where $\beta=1 (2)$ for systems with (without) time-reversal symmetry. Because
the short-time dynamics usually dominates the correlation functions, this
oversimplified approximation that doesn't consider individually the
contribution of other short trajectories is not expected to be very precise in
general. However, what is interesting about Eq.(\ref{kbmin}) is that it leads
to universal expressions for $C_\Omega$ and $C_R$, since as we will now see
the system-dependent features may be re-absorb by a rescaling of the
parameters. These universal results then serve as a guide (in particular in
experimental situations where the shape of the potential is not well known),
and are useful to obtain simple qualitative estimates of the different
characteristic scales involved.

Let's evaluate using (\ref{kbmin}) the variance of the fluctuations
of the grand potential and of the response function. Restricting
to zero temperature, we obtain from Eq.(\ref{co2}) 
\begin{equation} C_{\Omega}^0 =
C_{\Omega} (0,0,0) =  \frac{g_s^2}{2 \beta\pi^2}\frac{\hbar^2}{\tau_{min}^2}
 \ ,
\label{co0}
\end{equation} 
where $\tau_{min}$ is computed at $(x_0,\mu_0)$. This simple expression
relates the typical fluctuations to the period of the shortest orbit. For a
billiard, $\tau_{min} = (m/2\mu_0)^{1/2} L_{min}$, with $L_{min}$ the length
of the orbit. Weyl's law allows to express $\mu_0$ in terms of the average
number of particles in the gas $n= \overline{\cal N} (\mu_0,x_0)$. We find
$C_\Omega^0 \sim n^{2/d}$, with $d$ the space dimensionality. This may be
compared to the variance of the sum of $m$ energy levels located in a small
energy window around $\mu_0$, which is universal and proportional to $m^2 \log
m$ \cite{bls1}. If a finite temperature is considered, we find that
$C_{\Omega}^0$ is damped by a factor $4 (\tau_{min}/\tau_c) \exp (-
\tau_{min}/\tau_c)$ when $\tau_c <<\tau_{min}$.

For the variance of the response function we must compute the variance of the
distribution of the $Q_p$'s. We assume for chaotic systems a Gaussian
distribution \cite{gsbswz,blm}. The variance is $\langle Q^2 \rangle = \alpha
\ \tau$. Explicit expressions for $\alpha$ valid for chaotic billiards were
given in \cite{ls}. Its value depends on energy, with a dependence $\sim
E^{3/2}$. These results as well as the approximation (\ref{kbmin}) in
Eq.(\ref{cr2}) lead to
\begin{equation}
C_R^0 = C_R (0,0,0) = \frac{g_s^2}{\beta \pi^2} \frac{\alpha}{\tau_{min}} \ .
\label{cr0}
\end{equation}
It is interesting to compare $C_R^0$ to the ``local'' variance of the response
of the individual single-particle levels $E_k$ located in a small energy
window around $(x_0,\mu_0)$, $\langle v^2 \rangle = \langle (\partial E_k
/\partial x)^2 \rangle = 2\alpha g_s^2/ \beta \tau_H$ \cite{efkamm,ls}
($\tau_H = h \bar{\rho}$ is the Heisenberg time, conjugate to
the mean level spacing). We have
\begin{equation} \label{v2}
C_R^0 = (1/2 \pi^2) (\tau_H/\tau_{min}) \langle v^2 \rangle \ .
\end{equation}
Because $\tau_H >> \tau_{min}$, the electron gas amplifies the local
fluctuations. For billiards the amplification factor is proportional to
$n^{(d-1)/d}$, while the overall dependence on $n$ is $C_R^0 \sim n^{4/d}$.

The autocorrelation functions are now evaluated in the approximation
(\ref{kbmin}). Setting $\mu=T=T_0=0$ in (\ref{co2}) and using $\langle \cos(Q
x/\hbar) \rangle = \exp (-\alpha \tau x^2 / 2 \hbar^2)$ by the Gaussian assumption
we get
\begin{equation}
\frac{C_{\Omega} (\xi,0,0)}{C_{\Omega}^0} =  (1 - \xi^2/2) e^{- \xi^2/2} + 
\xi^4 \Gamma (0, \xi^2/2)/4 \ , \label{couniv}
\end{equation}
where $\Gamma (a,z)=\int_z^{\infty} t^{a-1} ~ e^{-t} ~ dt$. We have rescaled
the parameter according to
\begin{equation}
\xi= \sqrt{\alpha \tau_{min}}~ x/\hbar =
\sqrt{C_R^0/2 C_{\Omega}^0} ~ x \ , \label{presc}
\end{equation}
which makes, at this level of approximation, $C_\Omega /C_\Omega^0$ a
universal rapidly-decreasing function. The typical parameter scale over which
$C_\Omega$ decorrelates is $\delta x = \hbar/\sqrt{\alpha \tau_{min}}$. This
is consistent with the previous assumptions made, e.g., $\delta x$ is
small in the semiclassical $\hbar \rightarrow 0$ limit. It is interesting to
compare $\delta x$ to the analogous local scale for a single-particle level
\cite{sa1}, $\delta x_L = 1/{\bar \rho} \sqrt{<v^2>}$. Since $\delta x = g_s
\sqrt{\tau_H/2 \beta \pi^2 \tau_{min}} \delta x_L$, the scale of the Fermi gas
is larger by a factor $\sim n^{(d-1)/2 d}$.

Expressed in terms of the same rescaled variable, the autocorrelation function
of the response is
\begin{equation}
\frac{C_R(\xi,0,0)}{C_R^0} = e^{- \xi^2/2} - 3 \ \xi^2 \Gamma(0,\xi^2/2)/2  \ .
\end{equation}
Analogous results are obtained when computing the autocorrelation in energy, 
$C_\Omega (0,\epsilon,0)$ and $C_R (0,\epsilon,0)$, were the rescaled
parameter is $\epsilon = \mu \tau_{min}/\hbar$.

In order to test numerically the goodness of these approximations we have
considered an electron gas in a chaotic cavity. The cavity is the
two-dimensional Lima\c con billiard \cite{robnik}, whose shape is defined by
the conformal transformation $w=z+b ~ z^2 +c ~ e^{i\delta} z^3$. For parameter
values around $b=c=0.2$ it was numerically shown that the dynamics is
dominated by chaotic trajectories \cite{robnik,bs}. We take $x=\delta$ as the
external parameter. Fig.~1 shows the numerical results compared to the
theoretical curves. The approximation (\ref{kbmin}) provides a good
qualitative description of the main features. The deviations observed may be
due to contributions of periodic orbits other than the shortest one.

Analogous results may be obtained for diffusive systems by using the
appropriate form factor in Eqs.(\ref{co2}) and (\ref{cr2}). We concentrate on 
a particular example, the
calculation of the autocorrelation of the persistent current of an electron
gas in a disordered ring threaded by an Aharonov-Bohm flux $\phi$.

The current at thermodynamic equilibrium is given by
$I=-c \frac{\partial \Omega}{\partial \phi}$.
Only the fluctuating part of $\Omega$ depends on $\phi$. We replace $S_p/\hbar
\rightarrow S_p/\hbar + 2 \pi w_p \phi/\phi_0$ in Eq.(\ref{grandosc}) to
include the flux, 
with $\phi_0 = h c/e$ the flux quantum and $w_p$ the winding number of the
periodic orbit $p$ around the ring. The current in the ring is 
\begin{equation}
I (\lambda) = \frac{4 \pi \hbar^2 c g_s}{\phi_0} \sum_p \widetilde{A}_p ~ w_p 
\sin \left ( S_p/\hbar + 2 \pi w_p \lambda \right ) \ ,
\end{equation}
where we have introduced $\lambda=\phi/\phi_0$. The autocorrelation of the
current $C_I (\lambda,\lambda_R) = \langle I(\lambda_R + \lambda) ~
I(\lambda_R) \rangle$ is computed in the diagonal approximation taking into
account the contribution of the primitive orbits and of their time reversal
partners. As before, it can be expressed in terms of the form factor, which
now takes into account the diffusive motion of the particle \cite{as,ais}. For
simplicity we restrict the calculation to a quasi one-dimensional ring and set
the temperature to zero. At $\phi=0$ we get the well known variance of the
persistent current $C_I^0$ \cite{crg,ais}. Normalizing $C_I$ with this
variance and averaging both with respect to the reference flux $\lambda_R$, we
arrive at
\begin{equation}
\frac{C_I (\lambda)}{C_I^0} = \frac{1}{\zeta (3)} \sum_{w = 1}^{\infty}
\frac{\cos (2 \pi w \lambda )}{w^3} \ , \label{fluxuniv}
\end{equation}
where $\zeta (s)$ is the Riemann zeta function.

Eq.(\ref{fluxuniv}) is a universal function, independent of the specific
aspects of the dynamics (i.e., independent of the elastic time between
collisions or the diffusion constant), but depends on the dimensionality of
the system. No rescaling of the parameter was needed. This interesting fact is
due to the geometry of the sample and the physical quantity considered. Orbits
with zero winding number do not contribute to the current because they have no
flux dependence, while those with $w \geq 1$ have a period $\tau >> \tau_e$,
the elastic time. As a consequence, all the non-universal shortest
orbits gave no contribution, and the autocorrelation is a universal function
(at fixed $d$).

In conclusion, we have studied the statistical properties of a phase-coherent
noninteracting confined electron gas in response to a parameter. The results
are applicable to any confined fermionic system. The autocorrelations
$C_\Omega$ and $C_R$ were computed explicitly. Their behavior is described by
the shortest classical periodic orbits, making them system-dependent with in
general no universal regime. When the form factor is simplified to take into
account only the shortest orbit, it is possible to remove the dependence on
specific properties of the system by a rescaling determined by $\tau_{min}$.
The typical fluctuations of the gas are also expressed by simple formulas in
terms of $\tau_{min}$. This is different from the rescaling of parametric
correlations of single-particle levels, determined by $\tau_H$. This latter
time is of secondary importance in our context, and long orbits represent a
small correction to our results. In the case of persistent currents in a
diffusive Aharonov-Bohm geometry, the autocorrelation is independent of the
parameters of the system, without any rescaling, due to the suppression by the
geometry of the short-time orbits. The analytical results have been verified
by numerical simulations on chaotic and regular cavities.

We are especially grateful to H. Bruus, C. H. Lewenkopf and E. R. Mucciolo
for providing the program for the numerical simulation of the Lima\c{c}on
billiard.

\begin{figure}
\begin{center}
\leavevmode
\epsfysize=3.8in
\epsfbox{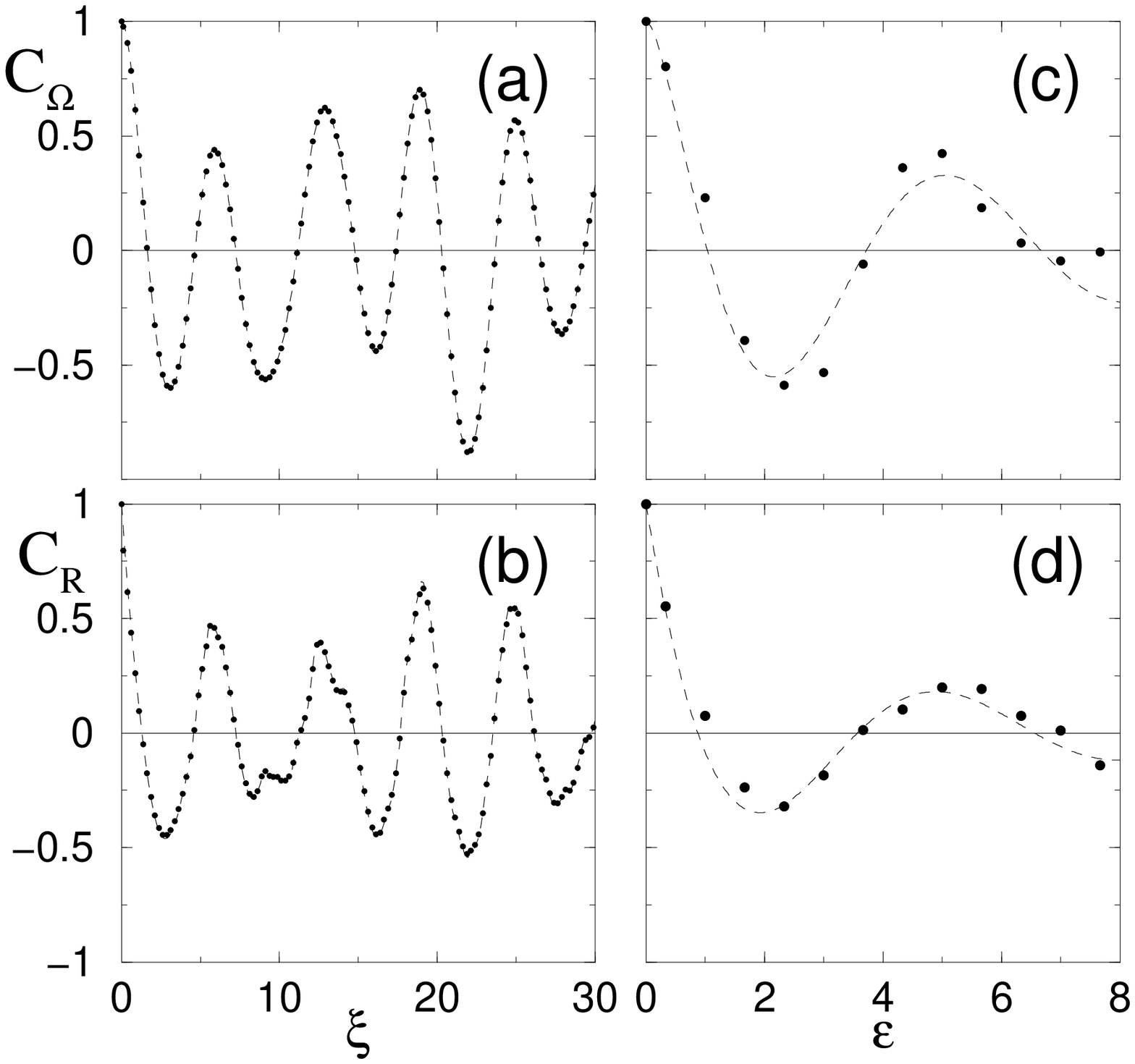}
\end{center}
\vspace{2.5cm}
\caption{Normalized autocorrelation functions (points) compared to theoretical
predictions (dashed lines): (a) $C_\Omega (\xi,0,0)/C_\Omega^0$ and (b) $C_R
(\xi,0,0)/C_R^0$ for the (integrable) rectangular billiard; (c) $C_\Omega
(0,\epsilon,0)/C_\Omega^0$ and (d) $C_R (0,\epsilon,0)/C_R^0$ for the chaotic
Lima\c{c}on billiard}
\label{fig}
\end{figure}

\end{document}